\documentclass[11pt,a4paper]{article}
\usepackage[utf8]{inputenc}
\usepackage[T1]{fontenc}
\usepackage{lmodern}
\usepackage{microtype}
\usepackage[margin=2.5cm]{geometry}
\usepackage{setspace}
\usepackage{parskip}
\usepackage{graphicx}
\usepackage{booktabs}
\usepackage{enumitem}
\setlist[itemize]{noitemsep}
\usepackage{multicol}

\usepackage{hyperref}
\hypersetup{colorlinks=true,linkcolor=blue,citecolor=blue,urlcolor=blue}
\usepackage{titling}
\usepackage{fancyhdr}
\usepackage{caption}
\usepackage{amsmath,amssymb}
\usepackage{aas_macros}
\usepackage{natbib}

\usepackage[printwatermark]{xwatermark}
\usepackage{xcolor}
\usepackage{graphicx}

\usepackage{titlesec}
\titlespacing*{\section}{0pt}{*2}{0.4\baselineskip}

\pretitle{\vspace*{-1cm}\begin{center}\LARGE\bfseries}
\posttitle{\par\end{center}\vskip 0.5em}
\preauthor{\begin{center}\large}
\postauthor{\end{center}}
\predate{\begin{center}\large}
\postdate{\end{center}}

\begin{document}

\begin{titlepage}
  \centering
  \vspace*{1cm}
  {\Huge\bfseries Expanding Horizons \\[6pt] \Large Transforming Astronomy in the 2040s \par}
  \vspace{1.5cm}

  {\LARGE \textbf{Disc Winds From Accreting Systems}\par}
  \vspace{1cm}

  \begin{tabular}{p{4.5cm}p{10cm}}
    \textbf{Scientific Categories:} & Time-domain; Accretion; Stars; Galaxies \\
    \\
    \textbf{Submitting Authors:} & Name: 
    Noel Castro Segura\\
    & Affiliation:  
    Dept. of Physics, University of Warwick (UK)\\
    & Email: 
    noel.castro-segura@warwick.ac.uk\\
    \\
    & Name: 
    Virginia Cúneo \\
    & Affiliation:  
    Leibniz-Institut für Astrophysik Potsdam (DE)\\
    & Email: 
    virginiacuneo@gmail.com\\
    \\
    \textbf{Contributing authors:} 
     & Francesco Tombesi (University of Rome “Tor Vergata”, IT),  
     Stefanie Fijma (ESO, API, NL),  
     Jesus Corral-Santana (ESO),  
     Alexandra Veledina (University of Turku, FI),  
     Alessandra Ambrifi (IAC, ES),  
     David Buckley (SAAO, ZA),
     Piergiorgio Casella (INAF-OAR, IT),
     Deanne L. Coppejans (University of Warwick, UK),
     Domitilla de Martino (INAF-OACN, IT),  
     Simone Scaringi (Durham University, UK).  
  \end{tabular}

  \vspace{1cm}

  \textbf{Abstract:}

  \vspace{0.5em}
  \begin{minipage}{0.9\textwidth}
    \small    
    What does the temporal evolution of disc winds tell us about accreting systems and the accretion process? Studies of accretion-disc outflows across all mass scales, including accreting white dwarfs, X-ray binaries, and active galactic nuclei, have shown that winds play a key role in regulating both the accretion flow and the surrounding environment. Disc winds therefore provide a common thread linking a broad range of scientific topics, from the microphysics of accretion to galaxy-scale feedback and evolution, as well as binary evolution and the predicted rates of energetic (multi-messenger) transient phenomena. Yet we still lack a comprehensive picture of the accretion-feedback process. Optical spectroscopy has revealed striking similarities across mass scales, hinting at common production mechanisms, and has shown that winds can evolve on timescales of only minutes. Progress, however, has been limited by their transient nature, sparse time coverage, and the lack of simultaneous, high-resolution spectroscopy. Time-domain facilities with high temporal and spectral resolution will allow us to track these events in high accretion-rate systems, constrain their launching mechanisms, and measure the mass, energy, and angular momentum they carry. This will provide crucial input for binary evolution models, wind feedback, and a unified view of accreting systems across mass scales.
  \end{minipage}

\end{titlepage}


\section{Introduction and Background}
\label{sec:intro}
Accretion, the process by which matter falls onto a central object, usually through a disc, is among the most fundamental phenomena in the universe. It powers sources across a wide range of scales, including  active galactic nuclei (AGN), X-ray binaries (XRBs; harbouring a stellar mass black hole or a neutron star), and accreting white dwarfs (AWDs). Accretion-disc winds are observed throughout this entire range of systems (e.g. [33, 7, 10] for AGN, XRBs and AWDs respectively). Studies of these outflows have reshaped our understanding of accretion by establishing disc winds as key regulators of the process (e.g. [11, 12, 26, 23, 18]). Yet, despite decades of observational and theoretical study, we still lack a complete understanding of the physical mechanisms that govern accretion and ejection onto compact objects. Closing this gap is essential because this process regulates the exchange of mass and energy between these systems and their environments, shaping binary evolution and driving galactic feedback. 

The smoking-gun signature of disc winds is blue-shifted absorption, often producing P-Cygni profiles; this spectroscopic footprint is observed from the X-ray to the infrared across a wide range of accreting systems, showing striking similarities that strongly suggest a common physical origin. Yet fundamental issues remain unresolved, including which mechanisms are responsible for driving disc winds, and whether those observed at different wavelengths share a common origin. Progress has been limited by the transient nature of winds, sparse time coverage, and the lack of simultaneous multiwavelength, high-resolution spectroscopy. Disc winds are thought to be driven by three main mechanisms: thermal, radiative, and magneto-centrifugal. Thermal winds arise when irradiation heats the disc surface so that gas becomes unbound at large radii [3], while radiative winds are powered directly by momentum transfer from the radiation field and thus depend on luminosity and ionisation state [6]. Magneto-centrifugal winds instead use large-scale magnetic fields anchored in the disc to accelerate gas along field lines [4, 28]. 

\textbf{Active galactic nuclei.} AGN host supermassive black holes (SMBHs) whose accretion discs launch powerful winds spanning a wide range of ionisation states and velocities—from UV/X-ray warm absorbers and broad absorption line winds to ultra-fast outflows reaching $\sim 0.1-0.3\,c$, seen in a substantial fraction of both local and high-z AGN [8, 15, 31, 32]. These absorption systems, often variable, span several orders of magnitude in velocity and ionisation parameter and are increasingly interpreted as different manifestations of a multi-scale and multi-phase stream carrying mass, momentum and kinetic power from sub-pc launching radii out to the kpc-scale interstellar medium [32,2]. These winds are now recognised as a primary channel for SMBH–galaxy coupling. Indeed, current estimates indicate that SMBH-driven winds can reach mass outflow rates comparable to or exceeding the accretion rate and kinetic powers of order a few per cent of the bolometric luminosity, possibly sufficient to sweep out the host gas reservoir, drive massive molecular outflows, and regulate both star formation and black hole growth [33, 9].

\textbf{X-ray binaries.} Early X-ray studies revealed highly ionized, equatorial winds (seen as blue-shifted Fe K absorption lines) that appear predominantly in the soft, disc-dominated accretion state and vanish in the hard, jet-dominated state (e.g. [34, 26]). 
Over the past decade, a growing number of detections have revealed that cool disc winds (e.g. [25]), traced by optical recombination lines, are present even during the hard accretion state—challenging the long-held view that winds are only present during soft states, as predicted by the thermal-driving scenario. Unlike the hot, highly ionised winds commonly seen in X-rays during soft states, these optical winds appear denser and slower, implying high mass-loss rates that may influence the accretion process shaping the evolution and duration of outbursts [30, 5]. Their relationship to the X-ray winds remains uncertain, but recent multiwavelength campaigns suggest they may represent different ionisation phases of the same outflow [7, 22]. While others propose that these cool winds are present throughout the outburst, with their observable properties modulated only slightly by the changing ionising spectrum [27]. Taken together, these findings paint a complex yet still incomplete picture of XRB outflows. Characterising the properties and geometry of disc winds is crucial not only for understanding how they couple to and reshape their surroundings, but also because, if magneto-centrifugal winds are dynamically important, current evolutionary models may underestimate angular-momentum losses by factors of $\gtrsim 40$ [14]. 

\textbf{Accreting white dwarfs.} AWDs have long been recognised as strong ultraviolet (UV) emitters, capable of powering winds through the line-driven mechanism, best traced via UV resonance lines (e.g. [13]). Interestingly, optical wind features closely resembling those seen in XRBs, have been identified in some AWDs that also show strong UV winds (e.g. [10]), pointing at common production mechanisms across accreting systems and wavelengths. Modelling tools for spectral signatures of winds were initially developed to interpret UV observations (e.g. [19, 17]); in recent years, however, they have improved substantially, including extensions to the optical band, as demonstrated by the \texttt{SIROCCO} ionization and radiative transfer code [20]. As a consequence, a series of recent results has shown that this is a rapidly developing field with strong potential to deliver tighter constraints on accretion physics. In particular, disc winds have provided a compelling interpretation of the observed spectra of a dwarf nova in outburst [29]; a diagnostic toolkit has refined observational wind identification through analysis of H$\alpha$ line profiles [36]; and clumping has emerged as a key ingredient in producing line-driven winds [21]. 

Altogether, this progress indicates that a coherent picture is emerging, and that further progress will be driven by observational tests capable of validating wind-launching scenarios.

\vspace*{-2mm}
\subsection*{Motivation and vision for the 2040s}
\vspace*{-2mm}
Observational advances during the last decade have revealed the intrinsically multi-phase nature of accretion disc winds, from discs around white dwarfs, neutron stars, stellar and supermassive black holes, largely independent of the accretor mass. However, efforts to characterise disc winds are still constrained by their intrinsically transient character, the patchy time coverage of current datasets, and the lack of coordinated, multiwavelength high-resolution spectroscopy. In the 2030s, coordinated multiwavelength campaigns using next-generation facilities will provide the key step forward: sensitive, high-resolution X-ray spectroscopy to characterise the hottest (and fastest components); optical/NIR spectroscopy and integral field spectroscopy to probe the cooler, denser phases; and (for AGN) sub-mm interferometry to map molecular outflows on galactic scales. Joint monitoring of systems across a wide spectrum of masses and system parameters, will refine key disc wind properties by comparing observables against increasingly sophisticated thermal, radiative and magnetically driven models. These efforts will likely establish the first scaling relations for parameters across the full mass range of accreting objects. Towards the 2040s, multi-messenger astronomy will open a new window on the Universe, providing a benchmark for population-level inferences on angular-momentum losses in binaries that will test our knowledge.

At the same time, these advances will expose the limitations of current facilities; short-lived wind episodes/signatures, rapid changes in ionisation state and geometry, and small-scale clumpiness will remain poorly constrained without instrumentation that combines high spectral resolution, true photon-counting time resolution, and flexible scheduling. In the 2040s, a facility with rapid response and high-cadence monitoring capability, spanning from the optical to NIR coupled with time-domain surveys and X-ray observatories, would allow us to capture disc winds as they form and evolve during dramatic state changes, apply time-dependent photoionisation diagnostics to measure densities directly (e.g. [16]), leveraging photon counting capabilities to determine the wind geometry (e.g. [7]), and use precise timing to perform reverberation-style lag studies between continuum and line emission to determine the launching radius. Polarimetric modes would add a powerful handle on ionisation, opening angles and clumpiness, by tying polarization degree and angle to the density and geometry (e.g. [35, 24]). Together, these capabilities, combined with the state-of-the-art radiation-hydrodynamic simulations, will finally allow us pin down the dominant launching mechanisms, quantify angular-momentum losses, and connect disc-scale physics to binary evolution and galaxy-scale feedback within a unified framework.

\vspace*{-2mm}
\section{Open Science Questions in the 2040s}
\label{sec:openquestions}
{\bf How are disc winds launched across all mass scales, and how are they connected?} What is the dominant driving mechanism as a function of system  parameters? Do disc winds observed at different wavelengths share a common origin? How do the physical properties and geometry of winds compare across system parameters? Addressing these questions will advance our fundamental understanding of accretion and accretion-coupled outflows across all scales.

{\bf What is the accretion-ejection balance of accreting systems at different scales?} How much mass and energy do disc winds carry, what is the impact of these outflows on their local environment, and how do winds regulate or interact with relativistic jets? Clarifying the accretion-ejection balance is essential for understanding galaxy-scale feedback and the interaction of these accreting systems with the interstellar medium, including their influence on star formation. 

{\bf How much angular momentum do disc winds remove from accreting systems?} How do these winds affect binary evolution across different mass scales, and do they regulate the efficient growth of compact objects? Quantifying wind-driven angular-momentum losses is crucial for incorporating physically motivated wind physics into binary evolution models and, in turn, for improving predictions of energetic (multi-messenger) transient rates.

\vspace*{-2mm}
\section{Technology and Data Handling Requirements}
\label{sec:tech}
\begin{itemize}\setlength{\parskip}{2pt} 
    \item Flexible, automated scheduling enabling immediate API/broker-based triggers for key transient monitoring.
    \item Mid- to High-resolution ($R\simeq 10,000-30,000$) optical-MIR ($0.4 - 2.4\,\mathrm{\mu m}$) on $4-20$\,m equivalent apertures (or more), ideally with polarimetry modes.
    \item Time-tagged, zero-readout-noise, energy-resolved photon counting for sub-second sampling.
    \item (Sub-)millisecond absolute timing calibration to allow cross correlations with other facilities at different wavelengths.
    \item Community led time allocation with disruptive rapid response mode for time-critical events and special categories for transient and multiwavelength programs.
    \item Automated data reduction pipelines delivering 1-D spectra or time-tagged event files for real-time decisions.
\end{itemize}

\vspace{-0.6\baselineskip}
\section*{References}
\vspace{-0.4\baselineskip}
\footnotesize
[1]~Ambrifi A. + 2025, A\&A, 694, A10; 
[2]~Arav N. + 2018, ApJ, 857, 6; 
[3]~Begelman M. C. + 1983, ApJ, 271, 7; 
[4]~Blandford R. D. + 1982, MNRAS, 199, 88; 
[5]~Casares J. + 2019, MNRAS, 488, 135; 
[6]~Castor J. I. + 1975, ApJ, 195, 15; 
[7]~Castro Segura N. + 2022, Nature, 603, 5; 
[8]~Chartas G. + 2002, ApJ, 579, 16; 
[9]~Cicone C. + 2018, ApJ, 863, 14; 
[10]~C\'uneo V. A. + 2023, A\&A, 679, A8; 
[11]~Done C. + 2007, A\&A Rev., 15;  
[12]~Fabian A. C., 2012, ARA\&A, 50, 45; 
[13]~Froning C. S., 2005, PASP, 330, 81; 
[14]~Gallegos-Garcia M. + 2024, ApJ, 973, 16; 
[15]~Ganguly R. + 2008, ApJ, 672, 10; 
[16]~Kosec P. + 2024, ApJ, 972, 3; 
[17]~Kusterer D. J. + 2014, A\&A, 561, A1; 
[18]~Laha S. + 2021, Nature Astronomy, 5, 1; 
[19]~Long K. S. + 2002, ApJ, 579, 72; 
[20]~Matthews J. H. + 2025, MNRAS, 536, 87; 
[21]~Mosallanezhad A. + 2025, arXiv:2512.0502; 
[22]~Mu\~noz-Darias T. + 2022, A\&A, 664, A10; 
[23]~Mu\~noz-Darias T. + 2016, Nature, 534, 7; 
[24]~Nitindala A. P. + 2025, A\&A, 694, A23; 
[25]~Panizo-Espinar G. + 2022, A\&A, 664, A10; 
[26]~Ponti G. + 2012, MNRAS, 422, L1; 
[27]~S\'anchez-Sierras J. + 2020, A\&A, 640, L; 
[28]~Scepi N. + 2018, A\&A, 609, A7; 
[29]~Tampo Y. + 2024, MNRAS, 532, 119; 
[30]~Tetarenko B. E. + 2018, Nature, 554, 6; 
[31]~Tombesi F. + 2010, A\&A, 521, A5; 
[32]~Tombesi F. + 2013, MNRAS, 430, 110; 
[33]~Tombesi F. + 2015, Nature, 519, 43; 
[34]~Ueda Y. + 1998, ApJ, 492, 78; 
[35]~Veledina A. + 2019, A\&A, 623, A7; 
[36]~Wallis A. G. + 2025, MNRAS, 543, 146.

\end{document}